\documentclass[prl,showpacs,showkeys,nofootinbib,twocolumn]{revtex4}

\usepackage{graphicx}  %
\usepackage{bm}  %

\newcommand{\simge}{\hspace*{0.2em}\raisebox{0.5ex}{$>$}
     \hspace{-0.8em}\raisebox{-0.3em}{$\sim$}\hspace*{0.2em}}

\newcommand{\beq}{\begin{equation}}
\newcommand{\eeq}{\end{equation}}
\newcommand{\bqa}{\begin{eqnarray}}
\newcommand{\eqa}{\end{eqnarray}}

\def\mqo2{{\!\!\!}}

\begin{document}

\title{Efimov States in a Bose-Einstein Condensate near a Feshbach Resonance}

\author{Eric Braaten}
\author{H.-W. Hammer}
\author{M. Kusunoki}

\affiliation{Department of Physics,
         The Ohio State University, Columbus, OH\ 43210, USA}

\date{April, 2003}

\begin{abstract}
Recent experiments with Bose-Einstein condensates of $^{85}$Rb atoms near
a Feshbach resonance have produced evidence for a condensate of diatomic
molecules coexisting with the atom condensate. It should also be possible
to create condensates of the triatomic molecules predicted by Efimov
coexisting with the atom and dimer condensates. The smoking gun for the
trimer condensate would be oscillatory dependence of observables on the
binding energy of the trimer. It may also be possible to deduce the
existence of the trimer condensate from the spectra of the bursts of atoms
and dimers created in the disappearance of the trimers.
\end{abstract}

\smallskip
\pacs{03.75.Fi, 03.65.Ge, 21.45.+v}
\keywords{Efimov states, Bose-Einstein condensates, Feshbach resonances}
\maketitle

In 1925, Albert Einstein predicted that at sufficiently low temperature,
an ideal gas consisting of noninteracting point-like identical bosons will
undergo a phase transition to a Bose-Einstein condensate (BEC), in which all
the particles are in the same quantum state \cite{Eins25}.
The superfluid phase of liquid $^4$He,
was for a long time the only physical system that could be
interpreted as a BEC.  The interpretation is
complicated not only by the strong interactions between the
atoms, but also by the high density of liquid $^4$He.
In 1995, Bose-Einstein condensation was observed in dilute trapped gases
of $^{85}$Rb, $^7$Li, and $^{23}$Na \cite{Ande95}.
These systems are close to the ideal gas limit considered by Einstein.
Although the interactions between the atoms give rise to large mean-field 
effects, they are still weak in the sense
that quantum fluctuations around the mean field are small.
Their strength is controlled by the diluteness variable $na^3$,
where $n$ is the number density and $a$ is the $s$-wave scattering length.
Moreover, the atoms can be accurately approximated
by point particles.  This approximation is controlled by the diluteness 
variable $n \ell^3$, where $\ell=(mC_6/\hbar^2)^{1/4}$, $C_6$
is the van der Waals coefficient, and $m$ is the mass of the atoms.

Atoms can have a large scattering length $a \gg \ell$ if a diatomic
molecule close to the 2-atom threshold creates a resonant interaction
between pairs of atoms. In this case, there is the possibility of a
qualitatively different BEC with pointlike
constituents ($n \ell^3 \ll 1$), but strong resonant 2-body interactions
($na^3\, {\rm \simge}\, 1$). We will refer to such a system as a
{\it resonant BEC}.
In contrast to liquid $^4$He, the pointlike nature of the
constituents in the resonant BEC provides hope that the behavior
of the many-body system is determined by
aspects of the few-body system that are dominated by the resonance.

The few-body problem for resonant atoms exhibits universal properties that
do not depend on the details of the interactions responsible for
the large scattering length. The most familiar of these properties in the
case $a>0$ is the existence of a shallow 2-body bound state 
(hereafter called simply the {\it dimer}) whose binding
energy in the limit $a \gg l$ is given by
\begin{equation}
|E_2| = \hbar^2/(ma^2)\,.
\label{B2}
\end{equation}
One particularly remarkable aspect of the
3-body system of resonant atoms is an accumulation of 3-body bound states
(so called {\it Efimov states}) near threshold as $a \to \infty$ \cite{Efi71}. 
As $a$ increases, new Efimov states appear below
the atom-dimer threshold at values of the scattering length that differ
by multiples of 22.7: $a_*$, $22.7a_*$, $515a_*$, $\ldots\,$. In the limit
$a=\infty$, there are infinitely many Efimov states with binding energies
that differ by multiples of 515. There is convincing
theoretical evidence that the ground state and excited state of the $^4$He
trimer are Efimov states \cite{BrH02}. Efimov states can be responsible
for dramatic energy dependence of scattering observables in the 3-body
sector. An interesting and challenging problem is to determine what effects
they would have in a resonant BEC. 
The possibilities include self-trapped atom condensates \cite{Bul99}
and mutually trapped atom and dimer condensates \cite{Cusack}.

A convenient system for studying the resonant BEC
is trapped atoms near a Feshbach resonance
\cite{Feshth}.
The scattering length $a$ is controlled by a magnetic field $B$ that can
be tuned to arbitrarily large values. 
Studies of $^{23}$Na condensates 
revealed large losses of atoms as $B$ was ramped 
towards or through the resonance where $a \to \pm \infty$ \cite{Sten99}.
There have also been experiments on
$^{85}$Rb condensates near the Feshbach resonance at 155 G \cite{Robe00}.
The most recent experiment has provided evidence for a molecular
condensate coexisting with the atom condensate \cite{Don02}.
The molecular condensate consists of the dimers
associated with the Feshbach resonance. The experiment also
revealed other surprising features involving a burst of high energy atoms
from the condensate and missing atoms.

In this Letter, we point out that it should also be possible in these
experiments to create coexisting condensates of atoms, dimers, and
Efimov trimers. We first summarize the relevant aspects of the
recent $^{85}$Rb experiment.
We then describe how a condensate of Efimov trimers could be
observed in similar experiments either through oscillations of
observables with a frequency determined by the trimer binding energy
or by the spectrum of the burst of atoms
emitted by the trimer condensate.

The experiments of Ref.~\cite{Don02} began with condensates
of $N\approx 16,000$ atoms at a magnetic field corresponding to a
small scattering length $a_{\rm init} \simeq 10 a_0$,
where $a_0$ is the Bohr radius. The condensate was subjected
to a time-dependent magnetic field $B(t)$, which produced a
scattering length $a(t)$ with the following time sequence:
1.~a first pulse up to a large scattering
length $a_{\rm max}\approx 2500 a_0$ and then down to an
intermediate value $a_{\rm hold}$, 
2.~a variable holding time $\tau$ with constant $a=a_{\rm hold}$, 
3.~a second pulse up to $a_{\rm max}$
and then down to a small scattering length $a_{\rm final}$, 
4.~a slow ramp up to a large scattering length $\simeq 1400 a_0$
followed by a holding time to allow the condensate to expand, 
5.~the sudden turn off of the magnetic field,
which increased the scattering length through the resonance $a = \pm\infty$
to the zero-field value $-480 a_0$.
The atoms were subsequently measured with a destructive
imaging technique sensitive only to atoms with energies less than
about 2000 nK. 
The time scales were 10's of $\mu$s for steps 1-3 and 5
and 1000's of $\mu$s for step 4.
The observed atoms consisted of two components: 
(1) a {\it remnant condensate} of $N_{\rm remnant}$ atoms 
with energies less than 3 nK, and 
(2) a {\it burst} of $N_{\rm burst}$ atoms 
that were emitted from the condensate near the time of steps 1-3
and had a typical energy scale of 150 nK.
Since these two numbers did not add up to $N$, there was also a 
third component: 
(3) $N_{\rm missing}$  {\it missing atoms}. 

The numbers $N_{\rm remnant}$ and $N_{\rm burst}$ were observed to be
periodic functions of $\tau$ with frequency 
$\nu_2=|E_2^{\rm hold}|/2\pi \hbar$,
where $|E_2^{\rm hold}|$ is the dimer binding energy
when the scattering length is $a_{\rm hold}$.
This suggests that the system during step 2
contained coexisting atom and dimer condensates. The
first pulse of $a(t)$ must have created the dimer condensate by a coherent
transfer of atoms from the atom condensate.
During the hold time, the two condensates would
have evolved by phases that differ by the angle $2 \pi \nu_2 \tau$.
The second pulse would have induced coherent transfers of atoms
between the two condensates. 
Atoms transferred from the dimer condensate 
could interfere with those in the preexisting atom condensate,
resulting in an oscillatory dependence of $N_{\rm remnant}$ on $\tau$
with frequency $\nu_2$.
Assuming the mechanism for the burst involves the atom condensates, 
$N_{\rm burst}$ should also oscillate as a function of $\tau$
with the same frequency $\nu_2$.
One surprising result from the experiment is that the phase of
$N_{\rm burst}$ is shifted relative to that of $N_{\rm remnant}$
by an angle that
depends sensitively only on the fall time of the pulse in step 3.
This feature seems to have been reproduced by the resonance 
field theory of Ref.~\cite{Holland}.
 
A time-dependent scattering length also allows atoms to be 
transfered from the condensates to incoherent states of energetic atoms.
We assume the evolution of $a(t)$ in step 4 was sufficiently
adiabatic that very few atoms were transferred out of the condensates.
We interpret the burst atoms and the missing atoms as the energetic atoms 
created by the changing scattering length during steps 1,3 
and during step 5, repectively.
The missing atoms would include atoms from the dissociation 
of the dimers in a dimer condensate due to the sweep through the 
Feshbach resonance in step 5.

\begin{figure}[tb]
\centerline{\includegraphics*[width=9cm,angle=0,clip=true]{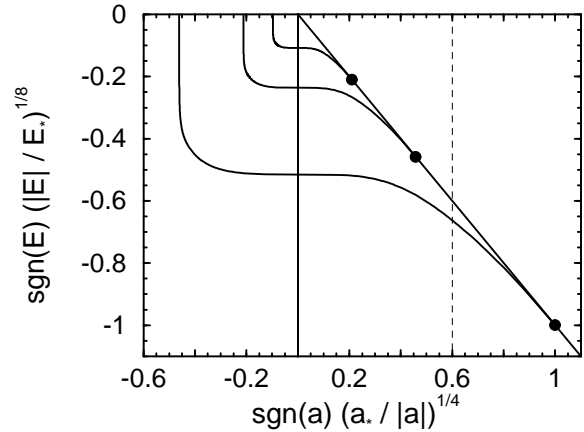}}
\vspace*{-1.0cm}
\caption{
The spectrum of Efimov states as a function of the scattering length
$a$ ($E_* \equiv \hbar^2/(ma_*^2)$).  
The diagonal solid line is the atom-dimer threshold.
The other solid lines are Efimov states.
The vertical dashed line shows a typical $a > a_*$.}
\label{fig:E3}
\end{figure}

The pulses that transfer atoms coherently to a dimer condensate 
could also transfer them coherently to a condensate of Efimov trimers.
The existence of the trimer condensate could be verified by observing 
oscillatory effects involving the trimer binding energy $|E_3|$. 
When the scattering length $a$ is large, Efimov theory predicts $|E_3|$ 
in terms of $a$ and a single 3-body parameter that is insensitive 
to variations in $a$ \cite{Efi71,BrH02,BHK99}. 
A convenient choice for this 3-body parameter is the critical value
$a_*$ of the scattering length for which an Efimov state is at the atom-dimer
threshold.  We measure energies relative to the 
3-atom threshold, so the atom-dimer threshold is  $|E_3|=|E_2|$. 
The parameter $a_*$ is related to the 3-body parameter 
$\Lambda_*$ introduced in Ref.~\cite{BHK99} by 
$s_0 \ln(a_* \Lambda_*) = 1.44\;{\rm mod}\;\pi$,
where $s_0=1.00624$. 
The Efimov spectrum is shown in Fig.~\ref{fig:E3}. 
If the scattering length is further increased to $22.7 a_*$, 
the binding energy of the first Efimov trimer becomes $|E_3|=6.8 |E_2|$
and a second Efimov trimer appears at the atom-dimer threshold
$|E_3|= |E_2|$.
Efimov theory predicts that if $a$ is large, 
the next critical value $a_*$ lies between $a$ and $22.7 a$.   
For $a>0$, large scattering length can
be defined by the condition that $|E_2|$ be given by (\ref{B2}).
For the experiment of Ref.~\cite{Don02}, this is satisfied to within 20\%
at $a=925 a_0$ ($B=158.5$ G) and it becomes increasingly accurate as $B$
approaches the resonance.  
Thus a reasonable range for the 3-body parameter is 
$925 a_0 < a_* < 21000 a_0$.
The value of $a_*$ depends exponentially on an angle that is 
determined by short-distance aspects of the 3-body problem. 
If we assume a uniform probability distribution for this angle, 
there is a 50\% probability that $a_* < 4400 a_0$.

If $a$ is close enough to $a_*$,
atom-dimer scattering will be dominated by the Efimov state near threshold.
The atom-dimer scattering length is \cite{BrH02,BHK99,BBH00}
\begin{equation}
a_{12} = [1.46 + 2.15 \cot (s_0 \ln (a/a_*) )]\,a\,.
\end{equation}
It diverges as $a \to a_*$, where the Efimov state is
at the atom-dimer threshold.  If $a$ is just above $a_*$,
we can approximate $E_3-E_2$ by $-\hbar^2/(2m_{12} a_{12}^2)$, where
$m_{12}=2m/3$ is the reduced mass of the atom and the dimer.
As $a \to a_*$, the difference in binding energies approaches
\begin{equation}
|E_3-E_2| \approx 0.16 \ln^2 (a/a_*)\; \hbar^2 /(ma^2) 
\hspace{0.5cm} (a>a_*)\,.
\label{B3-B2}
\end{equation}
This expression underestimates $|E_3-E_2|$ by 20\% at $a = 3.3 a_*$
and by a factor of 4 at $a = 22.7 a_*$.
A more accurate approximation can easily be obtained by the numerical
solution of the equation given in Ref.~\cite{BHK02}.
If $a$ decreases below $a_*$, the Efimov state disappears
from the spectrum.

If the number densities $n_1$ and $n_2$ of the atom and dimer condensates
are sufficiently large, it may be
necessary to take into account mean-field energy shifts.
The mean-field energy of an atom includes the terms
$2\pi\hbar^2 an_1/m$ and $3\pi\hbar^2 a_{12}n_2/2m$, while that of a
dimer includes a term $3\pi\hbar^2 a_{12}n_1/2m$.
The mean-field energy of the trimer and the term in the mean-field 
energy of a dimer proportional to $n_2$
are determined by solutions to the $n$-body problem with
$n \geq 4$ and are not known.

The simplest way to observe the trimer condensate is to choose
the magnetic field during the hold time $\tau$ so that
$a_{\rm hold} > a_*$, such as the dotted line in Fig.~\ref{fig:E3}.
In this case, there will be 3 coexisting
condensates during step 2. The first pulse will transfer atoms
coherently from the atom condensate into dimer and trimer condensates.
During the hold time, the dimer and trimer condensates
will accumulate phases relative to the atom condensate of
$2 \pi \nu_2 \tau$ and $2 \pi \nu_3 \tau$, respectively,
where $\nu_3 = |E_3^{\rm hold}|/2 \pi\hbar$
and $|E_3^{\rm hold}|$ is the trimer binding energy
when the scattering length is $a_{\rm hold}$.
The second pulse will transfer atoms
coherently between the three  condensates.
After the second pulse, which decreases the scattering length to
$a_{\rm final}< a_*$, the trimer condensate will have disappeared,
but it will have left behind a smoking gun. The numbers
$N_{\rm remnant}$, $N_{\rm burst}$ and $N_{\rm missing}$ will
exhibit oscillations in $\tau$ with frequency
$\nu_3-\nu_2$
on top of the oscillations with frequency $\nu_2$.

It could be difficult to increase $a_{\rm hold}$ above $a_*$,
because 3-body loss rates during the hold time increase like
$a_{\rm hold}^4$. It would be much easier to increase 
$a_{\rm max}$ above $a_*$.
If $a_{\rm hold}< a_* < a_{\rm max}$,
a trimer condensate could be created during the pulses.
But after each pulse, the trimer would be unstable and
would quickly decay into an energetic dimer and an atom.
This would destroy all information about the angle $2 \pi \nu_3 \tau$, 
so the numbers $N_{\rm remnant}$, $N_{\rm burst}$ and $N_{\rm missing}$ 
would oscillate only with frequency $\nu_2$.
The lowest oscillation frequency observed in the experiment of 
Ref.~\cite{Don02} was 10 kHz, which corresponds to 
$a_{\rm hold} \approx 2000 a_0$. 
If a second oscillation frequency in this data 
could be excluded, it would place an experimental lower bound 
$a_* > 2000 a_0$. 

It might be possible to find indirect
evidence for the trimer condensate by studying the
bursts of atoms created in experiments with a single pulse
that reaches a maximum $a_{\rm max}>a_*$, 
such as the dashed line in Fig.~\ref{fig:E3}.
The time sequence $a(t)$ would consist of a ramp from $a_{\rm init}$ 
to $a_{\rm max}$, a holding time, and then a ramp down to $a_{\rm final}$. 
The ramp up to
$a_{\rm max}$ would allow the coherent transfer of atoms from the atom
condensate into dimer and trimer condensates. The number of dimers and
trimers in these condensates would depend on the ramp rate, and could
therefore be varied. If the ramp down to $a_{\rm final}$ is sufficiently
rapid, it would create a burst of atoms with 2 components, 
one from the dimer condensate and one from the trimer condensates.

We first consider the burst of atoms emitted by
the dimer condensate.  If the scattering length is changed suddenly
from one large value $a$ to another $a'$, the probabilities
for the final states are determined by the overlaps
of the initial and final wavefunctions. The wavefunction for a dimer
at rest is proportional to $e^{-r/a}/r$. The wavefunction for
a scattering state consisting of two atoms with wavevectors
$\pm {\bf k}$ is $\cos({\bf k} \cdot {\bf r})
- (1/a + ik)^{-1} e^{ikr}/r$.
The probability that the dimer remains a dimer is
\begin{eqnarray}
P[D \to D] &=& 4a a'/ (a+a')^2 \hspace{1cm} a'>0,
\label{Pd}
\\
           &=& 0               \hspace{3cm} a'<0.
\nonumber
\end{eqnarray}
The probability density for the dimer to become a pair of atoms with
wavevectors $\pm{\bf k}$ is
\begin{equation}
\label{dPaa}
dP[D \to A\,A]
= {(a-a')^2 /(\pi^2 a^3 a'^2)
	\over (k^2 + 1/a^2)^2(k^2 + 1/a'^2)}\,d^3k\,.
\end{equation}
The distribution of the energy $E = \hbar^2 k^2/2m$ of the atoms
in the burst is obtained by multiplying (\ref{dPaa})
by $2N_ D$, where $N_D$ is the number of dimers in the dimer
condensate before the change in $a$.
The mean energy is $\langle E \rangle = (2a+|a'|)/(2|a'|) |E_2|$,
where $|E_2| = \hbar^2/(ma^2)$.

We now consider the burst emitted by a trimer condensate.
Determining the probabilities for the final states
in general requires solving a nontrivial 3-body problem.
However, as $a\to a_*$, the 3-body problem reduces to the 2-body problem
of an atom and dimer with resonant interactions. Therefore if $a$ and
$a'$ are both close to $a_*$, we can apply the analogs of (\ref{Pd}) and 
(\ref{dPaa}). The probability $P[T \to T]$ for the trimer
to remain a trimer is given by
(\ref{Pd}) with $a \to a_{12}$ and $a' \to a_{12}'$.
The probability density $dP[T \rightarrow A\,D]$
for a trimer to become an atom and a dimer with wavenumbers $\pm {\bf k}$ 
is given by (\ref{dPaa}) with $a \to a_{12}$ and $a' \to a_{12}'$.
The distribution of the energy
of the atoms in the burst is obtained by
multiplying by $N_T$, where $N_T$ is the number of trimers in the
trimer condensate before the change in $a$.
The mean energy of the atoms is 
$\langle E \rangle = (2a_{12}+|a_{12}'|)/(2|a_{12}'|) |E_3-E_2|$,
where $|E_3-E_2|$ is given in (\ref{B3-B2}). 
\begin{figure}[tb]
\centerline{\includegraphics*[width=9cm,angle=0,clip=true]{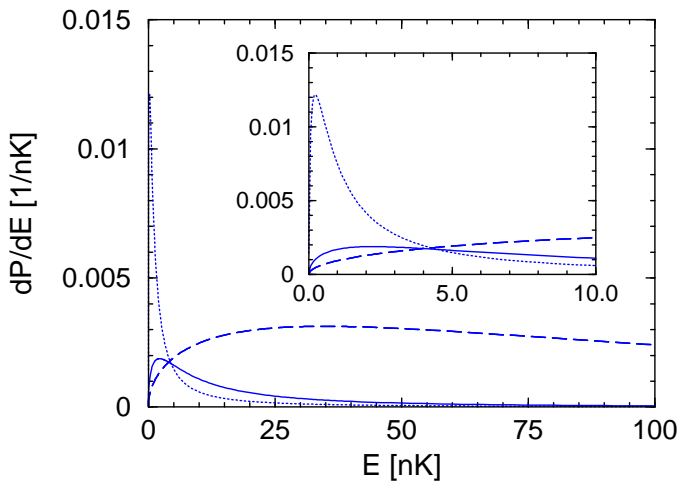}}
\vspace*{-0.0cm}
\caption{The energy distributions of the burst atoms
 emitted when the scattering length is suddenly changed from
 $a=3000 a_0$ to $1000 a_0$. The dashed line is the distribution of
 atoms emitted by the dimer condensate normalized to 1.
 The solid (dotted) line is the distribution of atoms 
 emitted by the trimer condensate for $a_*=1500 a_0$ ($2500a_0$)
 normalized to 1/25. The inset
 shows the low-energy region in more detail.}
\label{fig:dP}
\end{figure}
In Fig.~\ref{fig:dP}, we show the shapes of the energy distributions 
$dP/dE=(k/2E) dP/dk$ of the burst atoms
created if the scattering length is suddenly changed from
$a=3000 a_0$ to $1000 a_0$.
The correct normalizations are obtained by multiplying 
by $2 N_D$ for the dimer condensate 
and by $25 N_T$ for the trimer condensate.

If a method can be developed to image the dimers, it could
provide further evidence for a trimer condensate. If $a_{\rm max} < a_*$, 
most of the dimers should be in the dimer condensate.
However if $a_{\rm max} > a_*$, there should also be a burst of
dimers whose distribution in the energy $E = \hbar^2 k^2/4m$
is given by (\ref{dPaa}) with $a \to a_{12}$, $a' \to a_{12}'$,
and normalization $N_T$.
This burst of dimers would be another signature of the trimer condensate.

If the atom-atom scattering length $a$ is large, other 3-body observables
are completely determined by $a$ and the 3-body parameter $a_*$.
Once $a_*$ is determined, one may be able to test
the remarkable prediction of an approximate zero
in the 3-body recombination rate into an atom and the shallow 
dimer \cite{EGB99}.
The rate per unit time and volume of 3-body recombination events 
in a BEC is proportional to the cube of the number density:
$\nu = \alpha n^3/6$.
The contribution to $\alpha$ from recombination into the shallow dimer
is \cite{BBH00}
\begin{equation}
\label{alpha}
\alpha = 67.1 \sin^2[s_0 \ln (a/a_*) + 1.63] \;\hbar a^4/m\,,
\end{equation}
which vanishes at $a=0.20a_*,4.5a_*,\ldots$.
The contributions to $\alpha$ from recombination into deeply-bound
diatomic molecules also scale like $a^4$,
so the zeroes in (\ref{alpha}) should be interpreted
as local minima of $\alpha/a^4$.

There is a qualitative difference between the Efimov states
for heavy alkali atoms such as $^{85}$Rb
near a Feshbach resonance and those for $^4$He.
In the case of $^{85}$Rb, the 2-body bound states include
more deeply-bound diatomic molecules as well as the shallow dimer.
The Efimov states have a width, because they can decay into
an atom and deeply-bound molecules.  There are no reliable estimates
of the widths of the Efimov states.  These states are of physical
relevance only if their widths are small compared to their
binding energies.  The observation of a condensate of Efimov trimers
would provide dramatic confirmation of the importance of Efimov states
in the physics of cold atoms with resonant interactions.
It would also provide a glimpse into the rich and beautiful physics
of the resonant BEC.

This research was supported in part by DOE grant DE-FG02-91-ER4069 and
by NSF grant PHY-0098645. We thank Neil Claussen, Eric Cornell, Jason Ho, 
and Erich Mueller for useful discussions.

\end{document}